 \documentclass{aastex}

\def\gsim{\;\lower4pt\hbox{${\buildrel\displaystyle >\over\sim}$}\;}
\def\lsim{\;\lower4pt\hbox{${\buildrel\displaystyle <\over\sim}$}\;}

\shorttitle{}
\shortauthors{Fan \& Chiueh}

\begin{document}


\title{Determining the Geometry and the Cosmological\\ 
Parameters of the Universe through SZE Cluster Counts}


\author{Zuhui Fan\altaffilmark{1,2,3} and Tzihong Chiueh\altaffilmark{2,3}}


\altaffiltext{1}{Department of Astronomy and Astrophysics, The University
of Chicago, 5640 South Ellis Avenue, Chicago, IL 60637}
\altaffiltext{2}{Department of Physics, National Taiwan University,
1, Roosevelt Rd. Sec. 4, Taipei, Taiwan, R.O.C.}
\altaffiltext{3}{Institute of Astronomy \& Astrophysics, Academia Sinica,
P.O.Box 1-87, Nankang, Taipei, Taiwan 115, R.O.C.}


\begin{abstract}
We study Sunyaev-Zel'dovich Effect (SZE) cluster counts in
different cosmologies. It is found that even without the full
knowledge of the redshift distribution of SZE clusters,
one can still readily distinguish a flat universe with a cosmological
constant from an open universe. We divide clusters into a
low redshift group (with redshift $z\le 0.5$) and a high redshift
group (with $z\ge 1$), and compute the ratio of $r=N(z\le 0.5)
/N(z\ge 1)$, where $N(z\le 0.5)$ is the number of flux-limited ($S_{\nu}^{lim}$)
SZE clusters with $z\le 0.5$ and $N(z\ge 1)$ is the number
of flux-limited SZE clusters with $z\ge 1$. With about the same
total number of SZE clusters $N(z\ge 0)$, the $r$ value for
a flat universe with a non-zero cosmological constant and
that for an open universe occupy different regions in the
$S_{\nu}^{lim}-r$ plot for the most likely cosmological parameters
$0.25\le \Omega_0\le 0.35$ and $0.2\le \Gamma\le 0.3$,
where $\Omega_0$ is the matter density parameter of the universe, and
$\Gamma$ is the shape parameter of the power spectrum of linear density
fluctuations. Thus with a deep SZE cluster survey, the ratio $r$ can reveal,
independent of the normalization of the power spectrum, whether we
are living in a low-density flat universe or in an open universe.
Within the flat universe scenario, the SZE cluster-normalized
$\sigma_8$ is studied, where $\sigma_8$ is the r.m.s. density fluctuation
within the top-hat scale $8\hbox { Mpc}h^{-1}$ where $h$ is the Hubble constant
in units of $100\hbox { kms}^{-1}\hbox { Mpc}^{-1}$. A functional relation 
$\sigma_8\propto \Omega_0^{-0.13}$ is found. Combined with the 
X-ray cluster-normalized $\sigma_8\propto \Omega_0^{-0.52+0.13\Omega_0}$, 
one can put constraints on both $\Omega_0$ and $\sigma_8$ 
simultaneously.
\end{abstract}




\section{Introduction}

Clusters of galaxies are the largest virialized objects in the universe,
and contain valuable information of the universe and of the large-scale 
structure. There have been intensive studies on clusters from different
approaches, such as strong and weak gravitational lensing, X-ray, 
and optical observations. With the technical advents of interferometers,
the cluster's Sunyaev-Zel'dovich Effect (SZE) (Sunyaev \& Zel'dovich 1970,
1980; Birkinshaw 1999, Carlstrom et al. 1999), a spectral distorsion 
of Cosmic Microwave Background Radiation (CMB) due to scattering of CMB 
photons by hot electrons within clusters, has been becoming a new probe 
for the cluster study. As CMB photons pass through intracluster hot electrons, 
on average they gain energies through the inverse-Compton scattering, and as a 
result of this, the number of low energy photons decreases while the number 
of high energy ones increases. Thus for observations with the frequency
$\nu > (<) 219\hbox { GHz}$, hot clusters behave like emitting sources 
(absorbers) of photons. The equivalent temperature increment (or decrement) 
$\Delta T$ of CMB photons toward clusters is proportional to 
$\int n_e T_{gas} dl$, where $n_e$ is the number density of electrons, 
$T_{gas}$ is the hot gas temperature, and $dl$ is the line element along 
the line of sight. The integrated SZ effect of a cluster is then directly 
proportional to the cluster's gas mass if the gas is close to be isothermal.
Therefore the integrated SZ effect is not sensitive to the lumpy
structures of the gas and the gas fraction can be estimated relatively clean   
from the SZ effect in conjunction with lensing observations (e.g., Grego et al.
2000). On the other hand, because of the different dependence on $n_e$ of
the cluster's X-ray surface brightness ($S_x\propto \int n_e^2 \Lambda_{eH} dl$, 
where $\Lambda_{eH}$ is the X-ray cooling function) and of the SZ effect, 
the angular diameter distance to a cluster can be derived directly
from a joint analysis of the X-ray emission and the SZ effect through
modeling the gas density profile properly (e.g., Reese et al. 2000). 

Apart from the SZE studies for individual clusters, statistical investigations
on large number of SZE clusters can also yield very promising results. In fact 
several interferometric arrays have been proposed for surveying SZE clusters,
including the AMIBA (Array for MIcrowave BAckground) project which 
has been founded in Taiwan. Due to the frequency (energy) redshift 
dependence of the CMB photons, 
the SZ effect is independent of the redshift, which permits a SZE cluster 
survey to detect very high redshift clusters with relative ease. 
Therefore the cluster redshift evolution can be studied with high 
statistical significance. Another advantage of the SZE cluster studies over 
those of X-ray is that the integrated SZ effect of an individual cluster 
is directly proportional to the gas mass within the cluster (assuming the
gas is isothermal), which is in turn related to the total mass of the cluster, 
and the number of SZE clusters can be predicted analytically from the 
Press-Schechter formula (or other similar models) in a straightforward manner 
with certain qualifications. By contrast, because of the $n_e^2$ 
dependence of the X-ray surface brightness, the gas density profile
has to be modeled in the X-ray studies to estimate flux-limited
X-ray cluster number counts analytically, a procedure that can
introduce large uncertainties (the prediction on the number of 
temperature-limited X-ray clusters suffers less problem). 
One may investigate both aspects of clusters with numerical simulations,
but finite numerical resolutions and box sizes ultimately limit their 
applications, and the analytical analysis can be complementary.

Among other promising aspects, the redshift distribution 
of SZE clusters can be used to constrain cosmological parameters.  
The redshift distribution of SZE clusters with $\Omega_0=1$ is
distinctly different from those with $\Omega_0\sim 0.3$, and
the existence of several SZE clusters at redshift $z\gsim 1$ would strongly
exclude the $\Omega_0=1$ model. The difference between the redshift 
distribution of a low-density flat universe model 
and of a low-density open universe model is less dramatic, and
one needs a relatively large number of clusters at high redshift
to falsify them. The SZE cluster surveys are  
suitable for this purpose. By fully using the redshift distribution 
and the total number of SZE clusters, Haiman, Mohr, \& Holder (2000) 
studied constraints on the quintessence theory from future SZE (and X-ray) 
cluster surveys. The redshift of a cluster with $z\le 1$
can be obtained at least by using the photometric method around a
characteristic spectral break. For clusters with $z>1$, it however
appears difficult to measure the redshifts precisely
except for very large clusters.
In this paper, we propose a method which can distinguish,
within the parameter regime $0.25\le \Omega_0\le 0.35$ and 
$0.2\le \Gamma \le 0.3$,
a flat universe with a non-zero cosmological constant from an open
universe even without knowing the full redshift distribution 
of SZE clusters. We divide clusters into two groups: a low
redshift one with $z\le 0.5$, and a high redshift one with
$z\ge 1$. We study the ratio $r=N(z\le 0.5)/N(z\ge 1)$ for
different cosmologies with different parameters, where $N(z\le 0.5)$
is the total flux-limited number of clusters with $z\le 0.5$, 
and $N(z\ge 1)$ is the total flux-limited number of clusters with 
$z\ge 1$. It is found that $r$ can be used to disentangle a low-density 
flat universe from a low-density open universe. Notice that to compute $r$,
we only need to know the redshift range of a cluster (whether 
it is $z\le 0.5$ or $z\ge 1$) rather than its precise redshift.

We also investigate, within the framework of the flat universes with
a non-zero cosmological constant, the SZE cluster-normalized 
$\sigma_8$. The $\sigma_8-\Omega_0$ relation inferred from the 
SZE cluster counts is distinctly different from that from 
the X-ray cluster counts with the latter more steeper. This
difference can be used to limit the parameter regime of
$\Omega_0$ and $\sigma_8$ simultaneously.

The rest of the paper is organized as follows. Section 2 will present
the formulation for the study. The results will be shown in section 3.
Section 4 contains a summary.

\section{Formulation}

As CMB photons pass through a sea of hot electrons, 
their blackbody spectrum is distorted by the inverse Compton scattering. 
The SZ effect can be characterized by the Compton $y$ parameter, 
$$
y=\int n_e \sigma_T \bigg ({kT_{gas} \over m_e c^2}\bigg ) dl, \eqno (1)
$$
where $n_e$ is the number density of hot electrons, $\sigma_T=6.65\times
10^{-25} \hbox { cm}^2$ is the Thomson cross section, $k$ is the
Boltzmann's constant, $T_{gas}$ is the temperature of the hot intracluster gas,
$m_e$ is the electron mass, and $c$ is the speed of light.
The integration is along the line-of-sight, and $y$
parameter is proportional to the integrated thermal pressure along the
line-of-sight. When the electron temperature $T_{gas}$ is much 
higher than the temperature $T_{CMB}$ of the CMB photons, the CMB flux 
change due to the presence of a cluster can be written as
$$
S_{\nu}=S_{\nu}^{CMB} Q(x) Y,  \eqno (2)
$$
where $x=h_p \nu/kT_{CMB}$, $\nu$ is the frequency of the CMB photons,
$h_p$ is the Planck's constant, 
the unperturbed CMB flux $S_{\nu}^{CMB}=(2h_p \nu^3/c^2)/(e^x-1)$,
$$
Q(x)={x e^x \over e^x-1}\bigg [ {x\over \tanh(x/2)}-4\bigg ], \eqno (3)
$$
and  
$$
Y=R_d^{-2} \int y dA, \eqno (4)
$$
where $R_d$ is the angular diameter distance of the cluster, and the 
integration is over the projected area of the cluster.
It is seen that when $\nu \approx 219 \hbox { GHz}$, $Q(x)=0$. 
At the lower and higher frequency parts, $Q(x)<0$ 
and $Q(x)>0$, respectively. For AMIBA, $\nu =90 
\hbox { GHz}$, $x\approx 1.58$, and $Q(x)\approx -3.185$. 

We assume that the intracluster gas is isothermal and the 
gas mass fraction $f_{ICM}$ is a constant. Then we have (e.g., 
Eke, Cole, \& Frenk 1996)  
$$
Y={\sigma_T\over 2 m_e m_p c^2 } R_d^{-2} f_{ICM} (1+X) k T_{gas} M, 
\eqno (5)
$$
where $m_p$ is the proton mass, $X$ is the hydrogen mass fraction, and
$M$ is the total mass (including the dark matter) of the cluster. Here
we have used that the intracluster gas mass is dominated by hydrogen and 
helium. Further the gas is assumed to be in hydrostatic equilibrium with
the gravitational potential of the total mass of the cluster, then
$$
k T_{gas}=-{1\over [d \hbox {ln} \rho_{gas}(r)/ d\hbox {ln} r]_{r_{vir}}}
\mu m_p {GM\over r_{vir}}, \eqno (6)
$$
where $\rho_{gas}(r)$ is the radial density profile of the gas, 
$r_{vir}$ stands for the virial radius of the cluster, and 
$\mu=4/(5 X+3)$ is the mean molecular weight.
We have used, in equation (6), the virial mass to represent the total 
mass of the cluster. Let $\Delta_c$ be the average mass density
with respect to the critical density at redshift $z$ of the cluster formation,
then

\begin{eqnarray}
&&\qquad\qquad k T_{gas}=-{7.75 \over 0.5 [d \hbox {ln} \rho_{gas}(r)/ 
d\hbox {ln} r]_{r_{vir}}} \bigg ({ 6.8 \over 5 X+3 }\bigg )
\bigg ({M\over 10^{15} h^{-1} M_{\odot}}\bigg )^{2/3}
\nonumber \\ &&\qquad\qquad\qquad\qquad
\times (1+z) \bigg [{\Omega_0\over \Omega (z)}\bigg ]^{1/3}
\bigg ({\Delta_c \over 178}\bigg )^{1/3} \hbox { keV}, 
\qquad \qquad \qquad\qquad\qquad \qquad \qquad (7) \nonumber 
\end{eqnarray}
where $\Omega(z)$ is the density parameter at redshift $z$, and $h$
is the Hubble constant in units of $100 \hbox { kms}^{-1}
\hbox {Mpc}^{-1}$. Then 

\begin{eqnarray}
&&\qquad\qquad
S_{\nu}=2.29\times 10^4 {x^3\over e^x-1} Q(x)\times 1.70\times 10^{-2}
h\bigg ({f_{ICM}\over 0.1}\bigg ) \bigg ({1+X\over 1.76}\bigg ) 
\nonumber \\&&\qquad\qquad\qquad\quad
\times\bigg\{{7.75\over 0.5 [d\hbox {ln}\rho_{gas}(r)/d\hbox{ln}r]_{r_{vir}}}
\bigg \} \bigg ({6.8\over 5 X+3}\bigg ) \bigg ({R_d \over 100 h^{-1}
\hbox { Mpc}}\bigg )^{-2}
\nonumber \\&& \qquad\qquad\qquad\quad
\times (1+z)\bigg[{\Omega_0\over \Omega(z)}\bigg]^{1/3} 
\bigg ({\Delta_c \over 178}\bigg )^{1/3}\bigg ({M\over 10^{15} h^{-1}
\hbox{ M}_{\odot}}\bigg)^{5/3}\hbox{ mJy},\qquad\qquad\qquad\quad (8)
\nonumber 
\end{eqnarray}

where $1\hbox { mJy}=10^{-26} \hbox { ergcm}^{-2}\hbox {s}^{-1}
\hbox {Hz}^{-1}$. In the following we will use the value 
$[d \hbox {ln} \rho_{gas}(r)/ d\hbox {ln} r]_{r_{vir}}=2$, which is 
consistent with both the observational and the numerical simulation results. 
In fact, the results for different values of $[d \hbox {ln} \rho_{gas}(r)/
d\hbox {ln} r]_{r_{vir}}$ can be obtained from our analyses
by rescaling the overall flux $S_{\nu}$ up or down, as can be seen from equation (8).
The cosmology enters the relation (8) between $S_{\nu}$ and $M$ through the 
angular diameter distance $R_d$, the 
density parameters $\Omega_0$ and $\Omega(z)$, and the over density 
parameter $\Delta_c$. We will calculate the flux-limited SZE cluster counts. 
Note by using equation (8) to calculate $M_{lim}$ from a given flux limit 
$S_{\nu}^{lim}$, we have implicitly assumed that the counts are for 
unresolved clusters.
An array of interferometers must, however, have a minimum
baseline which is essentially limited by the dish diameter $D$. Signals
from angular scales larger than about $\lambda/(2D)$ are lost
where $\lambda$ is the observing wavelength. For AMIBA
there are two sets of dishes with $D=1.2 \hbox { m}$ and 
$0.3 \hbox { m}$, respectively. For $\nu=90 \hbox { GHz}$, 
$\lambda\approx 0.33 \hbox { cm}$,
and $\lambda/(2D)\sim 4.7\hbox { arcmin}$ for $D=1.2 \hbox { m}$ and 
$18.9\hbox { arcmin}$ for $D=0.3\hbox { m}$. For the smaller set of 
dishes, there should not be of any loss of signals from clusters. 
We have estimated the mass limit
for $D=1.2\hbox { m}$ by simply cutting off any signals from
$\theta \ge 4.7\hbox { arcmin}$. The hydrostatic equilibrium gas
density profile has been used by assuming that the underlying dark matter 
distribution has a universal density profile (Navarro, Frenk \& White 1997). 
The mass limit estimated here is not very different from
that for the unresolved clusters at $S_{\nu}^{lim} \sim 5 \hbox { mJy}$ 
of the AMIBA design. Thus our studies presented in the next section 
will only consider the unresolved cluster counts. Holder et al. 
(1999) determined $M_{lim}$ by performing mock observations 
appropriate for a proposed interferometric array which consists of 
ten $2.5\hbox { m}$ telescopes operating on $\nu=30\hbox { GHz}$ 
(Mohr et al. 1999) on simulated clusters;
the shape of the $M_{lim}(z)$ is similar to that for unresolved
clusters. In the work of Haiman et al. (2000), the mass limit
for their fiducial model is from Holder et al. (1999), and
$M_{lim}$ for other cosmological models is obtained
by using the same scaling relation as that of equation (8). They 
found that the mass limit from the scaling relation agrees with the
mock survey results better than $10\%$ for the two testing cosmological models. 
We use the Press-Schechter formalism (Press \& Schechter 1974) to 
calculate the number of SZE clusters. The comoving number density of 
clusters of mass $M$ with width $dM$ is,
$$
n(M)dM=\bigg ({2\over \pi}\bigg )^{1/2} {\rho_0\over M}
{\delta_c(z)\over \sigma_0^2} {d\sigma_0\over dM}
\exp\bigg (-{\delta_c^2(z)\over 2\sigma_0^2}\bigg ),
\eqno (9)
$$
where $\rho_0$ is the present mass density of the universe, 
$\delta_c(z)$ the linear overdensity threshold 
for collapse at redshift $z$, and $\sigma_0$ the r.m.s. linear density 
fluctuation on the scale corresponding to $M$. Notice that 
$\delta_c(z)$ and $\sigma_0$ are computed from the
extrapolated-to-present linear density perturbations.       
Then the differential number of SZE clusters is
$$
{dN\over dz d\Omega}={dV\over dz d\Omega}\int_{M_{lim}(z)}
n(M) dM, \eqno (10)
$$
where $d\Omega$ is the solid angle element, $dV$ is the comoving
volume element which is dependent of cosmologies, and
$M_{lim}(z)$ is calculated from equation (8).




\section{Analyses}

\subsection{ Flat Models versus Open Models}

For a cold-dark-matter universe, the power spectrum of the linear
density fluctuation field can be written as (e.g., Efstathiou, Bond, 
\& White 1992)
$$
P(k)={Bk^n\over \{1+[ak+(bk)^{3/2}+(ck)^{2}]^{\nu}\}^{2/\nu}}\quad , 
\eqno (11)
$$
where $a=(6.4/\Gamma)h^{-1}\hbox{ Mpc}$, $b=(3.0/\Gamma)h^{-1}\hbox{ Mpc}$,
$c=(1.7/\Gamma)h^{-1}\hbox{ Mpc}$, $\nu=1.13$, $\Gamma$ is the
shape parameter of the power spectrum, which is related to the time
of equal matter-radiation energy density in the universe,  
$h$ is the Hubble constant in units of $100\hbox{ kms}^{-1}\hbox{ Mpc}^{-1}$, 
$B$ represents the perturbation amplitude, and $n$ is the power index 
which is taken to be $n=1$. 

The object of this subsection is to distinguish the low-density flat
universes from the low-density open universes. Before proceeding to this issue,
let us first consider examples of the redshift distribution of SZE clusters
for some popular cosmologies. In Fig.1, the SZE clusters' redshift 
distribution is shown for (1). $\tau$-CDM (White, Gelmini, \& Silk
1995) (solid line): $\Omega_0=1$, $h=0.5$, $\Gamma=0.25$, and $\sigma_8=0.52$ 
where $\sigma_8$ is the r.m.s. density fluctuation within the top-hat 
scale $8\hbox { Mpc}h^{-1}$; (2). SCDM (dash-triple dotted line):
$\Omega_0=1$, $h=0.5$, $\Gamma=0.5$, and $\sigma_8=0.52$;
(3). open CDM (dash-dotted line): $\Omega_0=0.3$, $\Omega_{\Lambda}=0$, 
$h=0.83$, $\Gamma=0.25$, and $\sigma_8=0.87$; (4). $\Lambda$CDM
(dotted line): $\Omega_0=0.3$, $\Omega_{\Lambda}=0.7$,
$h=0.83$, $\Gamma=0.25$, $\sigma_8=0.93$. Here the normalization 
factor $\sigma_8$ is determined from the X-ray cluster counts
(Eke et al. 1996). It is clearly seen that the redshift 
distribution of $\Omega_0=1$ models is drastically different
from that of low-$\Omega$ models. Because of the continuing growth
of the linear density perturbation for the $\Omega_0=1$ models, their cluster 
numbers at high redshift (e.g., $z\ge 1$) are very tiny 
when the perturbations are normalized to the local cluster abundance. 
The presence of a few clusters at $z\ge 1$ would strongly falsify 
the $\Omega_0=1$ models. In fact, the existence of high redshift optical 
or X-ray  clusters has been used to constrain the range of $\Omega_0$ 
(e.g., Bahcall \& Fan 1998).
On the other hand, the differences of the redshift distribution between the two low-density cosmological models are not as impressive as the differences
between them and the $\Omega_0=1$ models, but are still rather substantial
at high redshifts. The number of SZE clusters drops rapidly for 
the $\Lambda$CMD model, while for the low-density open model its redshift 
distribution has a long tail at high redshifts. The different behaviors 
of the two models are mainly caused by the different angular diameter 
distances $R_d$ [see equation (8)]. 

In the following we will study the ratio of the low and high redshift SZE 
clusters for the two types of low-density cosmologies. 
Specifically, the ratio of $r=N(z\le 0.5)/N(z\ge 1)$ is considered. 
The relative number of high and low-redshift clusters with their masses
above a given threshold has been used in determining separately $\Omega_0$ 
and $\sigma_8$ by Fan, Bahcall and Cen (1997). Their study is different
from ours in several aspects, but both analyses take the
advantage of the dependence of the cluster evolution on cosmologies.

The model with $\Omega_0=0.3$, $\Omega_{\Lambda}=0.7$, 
$\Gamma=0.25$ and $\sigma_8=0.93$ is chosen to be the fiducial one. 
The particular $\sigma_8$ value is from the observed X-ray cluster counts.
For other models, the cosmological 
density parameter is taken to be $0.25\le \Omega_0\le 0.35$. 
We vary $\Gamma$ in the range of $0.2\le \Gamma\le 0.3$, consistent with 
large-scale structure studies (Peacock \& Dodds 1994, Dodelson \& Gaztanaga 
2000). 

We determine $\sigma_8$ of a specific model in such a way that it 
gives rise to about the same total number of flux-limited SZE clusters 
as that of the fiducial one at $S_{\nu}^{lim}\approx 6.2\hbox{ mJy}$. 
We will refer to this as SZE-cluster-normalized $\sigma_8$.
For the moment we pretend not to know the X-ray or the optical 
cluster normalization (except for the $\sigma_8$ of the fiducial 
model), and only the SZE cluster counts are used in determining 
both $\sigma_8$ and $r$. In fact the derived $\sigma_8$ from SZE 
cluster counts is consistent with that from X-ray or optical 
cluster counts for the parameter range we are considering. But 
later we will see that if we require the same total number of 
SZE clusters for two models with significantly different values 
of $\Omega_0$ (e.g., $\Omega_0=0.2$ versus $\Omega_0=0.4$), at 
least one of the two $\sigma_8$ will have to be out of the 
current X-ray cluster constraint. This inconsistency can be 
used in turn to falsify different models.  

In Fig.2 we show $r$ versus $S_{\nu}^{lim}$ for the fiducial cosmology
(solid line), and for low-density open cosmologies. To avoid crowding, for
open cosmologies only the 
highest and the lowest $r$ (dotted lines) for $0.25\le \Omega_0 \le 0.35$ and
$0.2\le \Gamma \le 0.3$ and the results for $\Omega_0=0.3$ with
$\Gamma=0.2, 0.25 \hbox { and } 0.3$ (dashed lines) are plotted. 
It is seen that 
the $r$ range for the open universes is quite separated from that of the
fiducial model. At $S_{\nu}^{lim}\approx 6.2 \hbox { mJy}$, 
$dN(z\le 0.5)/d\Omega\approx 4.0\hbox { deg}^{-2}$ and 
$dN(z\ge 1)/d\Omega\approx 0.22\hbox { deg}^{-2}$ for the low-density
flat fiducial model. For a SZE cluster survey which covers $50 \hbox { deg}^2$, $N(z\le 0.5)\approx 200$ and $N(z\ge 1) \approx 11$. Consider the 
Poisson noise, then $\sigma_r/r\approx (1/200+1/11)^{1/2}\approx 0.31$, 
where $\sigma_r$ is the standard deviation of $r$. Thus the largest 
$r$ for the set of open universes is about $2\sigma$ away from the
fiducial $r$. The open model with $\Omega_0=0.3$ and $\Gamma=0.25$
is about $3\sigma$ away. This demonstrates that if the $r$ value
and the total number of SZE clusters from observations 
are indeed around the values of the fiducial model,
low-density open models can be excluded at the $2-3\sigma$ level
depending on how well the $\Omega_0$ and $\Gamma$ parameters have been 
determined. With larger surveys, the exclusion can be made 
with higher statistical significance. We emphasize that different models
are SZE cluster normalized, and thus the above conclusion is independent
of the `real' normalization (where `real' normalization means the conventional
normalization from X-ray/optical cluster observations).
On the other hand, since the total number of 
SZE clusters is sensitive to the normalization factor, the combined
analysis on X-ray cluster counts and on SZE cluster counts could  
falsify cosmological models if the normalizations determined separately from 
the X-ray clusters and from the SZE clusters disagree.
This point will be elaborated in the next subsection.

Attentive readers may have suspected that if one decreases $\Omega_0$ 
and $\Gamma$ for a flat model, its $r$ range can get closer 
to those of open models. In Fig.3, we show the result for
$\Omega_0=0.25$, $\Omega_{\Lambda}=0.75$, $\Gamma=0.2$,
and $\sigma_8\approx 1.01$ (Model 1, solid line) along with some of the results
for open models: $\Omega_0=0.35$ and $\Gamma=0.3$ (Model 2, dotted line);
$\Omega_0=0.25$ and $\Gamma=0.3$ (Model 3, upper dashed line); $\Omega=0.25$ 
and  $\Gamma=0.25$ (Model 4, middle dashed line); $\Omega_0=0.25$ and 
$\Gamma=0.2$ (Model 5, lower dashed line). 
All other open models with $0.25\le \Omega_0 \le 0.35$ and 
$0.2\le \Gamma \le 0.3$ have the $r$ value in between the results of Model 
2 and 5. The open models have been normalized so that they contain about 
the same total number of SZE clusters as that of Model 1 at
$S_{\nu}^{lim}\approx 6.2 \hbox { mJy}$. At this $S_{\nu}^{lim}$, 
Model 1 has $r\approx 10.4$ and Model 2 has $r\approx 7.0$.  
For Model 1 $dN(z\le 0.5)/d\Omega\approx
3.6\hbox { deg}^{-2}$ and $dN(z\ge 1)/d\Omega\approx 0.35 \hbox { deg}^{-2}$, 
and the standard deviation of $r$ for a $50\hbox { deg}^2$ survey is then 
$\sigma_r\approx 2.6$. Thus Model 2 differs from Model 1
at about $1\sigma$ level. This difference itself may not be large
enough to distinguish the two models. We however notice that the two parameters
$\Omega_0$ and $\Gamma$ for the two models lie toward the opposite limit
of our considered range: $\Omega_0=0.25$ and $\Gamma=0.2$ for Model 1, 
and $\Omega_0=0.35$ and $\Gamma=0.3$ for Model 2. Hence if 
$\Omega_0$ or $\Gamma$ can be constrained to better degrees by other 
observations, the low-density flat model and open models can be distinguished
at a higher level of significance (for example, the Sloan Digital
Sky Survey would give a better constraint on the $\Gamma$ parameter). 
This can be seen from the $r$ differences at 
$S_{\nu}^{lim}\approx 6.2 \hbox { mJy}$ which are
at $1.9\sigma$ and $2.3\sigma$ levels between Model 1 and Model 3, 
and between Model 1 and Model 4, respectively. Most drastically
Model 1 and Model 5 have the same $\Omega_0$ and $\Gamma$, and the difference
in $r$ increases to the $3\sigma$ level.
  
We show in Fig.4 the $\Gamma$ dependence of $r$ for $\Omega_0=0.3$
and $\Omega_{\Lambda}=0.7$. The three curves are for $\Gamma=0.2,
0.25$ and $0.3$, respectively. All the models are normalized
to contain about the same SZE clusters as our fiducial model
at $S_{\nu}^{lim}\approx 6.2 \hbox { mJy}$. It is seen that 
$r$ is not very sensitive to $\Gamma$. In other words, 
$\Gamma$ cannot be strongly constrained by the SZE cluster
counts alone. It is interesting to note that if the three models
are all normalized to $\sigma_8=0.93$, the trend in $r$ is
reversed from that shown in Fig.4, i.e., $r$ with $\Gamma=0.2$ is the largest,
and $r$ for $\Gamma=0.3$ is the smallest. But the curves occupy
the same region as that in Fig.4.

To see the $\Omega_0$ dependence of $r$, we plot $r$ in Fig.5 for
low-density flat models with $\Omega_0=0.25, 0.3$ and $0.35$, respectively.
The $\Gamma$ parameter is taken to be $0.25$ for all the three models, 
and as before, they are normalized according to the total number of
SZE clusters of the fiducial model at $S_{\nu}^{lim}\approx 6.2
\hbox { mJy}$. With such normalizations, $r$ is not sensitive
to $\Omega_0$ either. 


We conclude that $r$ is a very useful
quantity to differentiate the low-density flat cosmological
models from the low-density open cosmological models. Within the likely
parameter regime of $\Omega_0$ and $\Gamma$, the quantity $r$ is not sensitive
to either of them if different models are normalized 
consistently to the SZE cluster counts. This method is independent
of the means by using the CMB measurement, and thus provides an important 
test on our understanding of the universe and of the formation of the 
large-scale structure.    

Our conclusion above is not sensitive to the specific total number density of
SZE clusters used to normalize different models. For instance, 
if we increase the total SZE cluster number density from $5.6\hbox { deg}^{-2}$
to $8.0 \hbox { deg}^{-2}$ (the corresponding change of $\sigma_8$ of the 
fiducial model is from $0.93$ to $1.0$), the $\Lambda$CDM model
with $\Omega_0=0.3$ and $\Gamma=0.25$ and the open
model with the same $\Omega_0$ and $\Gamma$ can also be distinguished  
at about $3\sigma$ level for a $50\hbox { deg}^2$ survey. 
The operational steps to apply our method to observations are
(1) to normalize different models to the observed total SZE cluster counts;
(2) to calculate the $r$ value for different models;
(3) to find the $r$ value from observations;
(4) to compare (3) with (2). 

On the other hand, the relatively large separation of the $r$ value
between the $\Lambda$CDM and the open CDM models shown above is 
restricted to our considered parameter intervals of $\Omega_0$ and
$\Gamma$. If the $\Omega_0$ range is increased to $0.2 \le \Omega_0 \le 0.4$, 
there will have some overlaps in $r$ for the two types of models. 
For example, the $r$ value of the $\Lambda$CDM model with $\Omega_0=0.2$
falls into about the same range as that of the open CDM model
with $\Omega_0=0.4$. Thus solely with the SZE total counts and the $r$
value, the two models cannot be differentiated clearly. However, 
one would find that in order to have the same total number of 
SZE clusters, at least one of the two $\sigma_8$ must be
out of the range allowed by the observed X-ray cluster counts. Therefore
an exclusion is possible based on both the SZE cluster observations
and the X-ray observations. Stated somewhat differently, 
in order to make a relatively clean distinction between the 
$\Lambda$CDM and the open CDM models 
by using $r$ alone, the parameters $\Omega_0$ and $\Gamma$ 
must be pre-determined by other observations to a relatively
fine degree. In conjunction with X-ray or optical cluster
observations, the SZE cluster counts can be used to 
distinguish the two types of models for wider parameter regimes. 

\subsection{Constraints on $\Omega_0$ and $\sigma_8$ for a Flat Universe }

Results from the recent Boomerang Cosmic Microwave Background (CMB)
radiation observation show the 
first Doppler peak at $l\approx 200$ beautifully (de Bernardis et al. 2000), 
and make the flat universe be widely accepted (Hu 2000, Tegmark \& 
Zaldarriaga 2000). If we indeed live in a flat universe with a 
non-zero cosmological constant, then, as we will describe below,
the combined analyses of X-ray cluster counts and SZE cluster counts can give 
rise to constraints on $\Omega_0$ and on $\sigma_8$ even without 
the cluster redshift information.

The key here is that the dependence of $\sigma_8$ on $\Omega_0$ inferred 
from X-ray cluster counts is different from that from SZE cluster counts.
The X-ray cluster counts yielded the relation 
$\sigma_8=(0.52\pm 0.04)\Omega_0^{-0.52+0.13\Omega_0}$ for $\Omega_0+
\Omega_{\Lambda}=1$ (Eke et al. 1996). Here we study the expected 
$\sigma_8-\Omega_0$ correlation from SZE cluster counts.  

The formulation presented in Sec. 2 are used to study the SZE  
$\sigma_8-\Omega_0$ relation. Or more clearly, the following assumptions
are employed: (1). The intracluster gas is in hydrostatic equilibrium
in the gravitational potential well of the total cluster mass;
(2). The gas is isothermal and the gas mass fraction is a constant
among different clusters; (3). The collapse is approximately spherical;
(4). The Press-Schechter formula is approximately correct
in predicting the number of clusters. We would like to point
out that same approximations have been used in deriving the
$\sigma_8-\Omega_0$ correlation from X-ray cluster observations
(Eke et al. 1996)

The model with $\Omega_0=0.3$, $\Omega_{\Lambda}=0.7$, $\Gamma=0.25$,
and $\sigma_8\approx 0.93$ is taken to be the fiducial one.
We first compute the total surface number density 
of SZE clusters at $S_{\nu}^{\lim}=6.2\hbox { mJy}$ for the
fiducial model, and then find $\sigma_8$ values for other
$\Lambda$CDM models with different $\Omega_0$ so that 
they have the same total surface number density of SZE clusters
at $S_{\nu}^{\lim}=6.2\hbox { mJy}$ as the fiducial model.

The results are shown in Fig. 6. It is found that the relation can be nearly
perfectly described by $\sigma_8=A_S\Omega_0^{-0.13}$ (the solid line
in Fig. 6), where $A_S$ is a numerical factor which is equal to $0.794$ 
in our analysis. The dependence of the SZE cluster-normalized $\sigma_8$ 
on $\Omega_0$ is much weaker than that of the X-ray cluster-normalized
$\sigma_8$, which can be understood from Eqns (7) and (8). The
X-ray cluster-normalized $\sigma_8$ is calculated from 
temperature-limited cluster counts (Eke et al. 1996).
The cluster mass limit $M_{lim}$ dervied from a given cluster gas 
temperature threshold is proportional to 
$[\Omega_0/\Omega(z)]^{-1/2}\Delta_c^{-1/2}$ 
[Eqn. (7)]. By contrast, given a SZ flux limit $S_{\nu}^{lim}$, we 
have $M_{lim}\propto [\Omega_0/\Omega(z)]^{-1/5}\Delta_c^{-1/5}$ [Eqn. (8)]. 
Since the cluster number counts from the Press-Schechter formalism is
sensitive to the mass limit, the inferred $\sigma_8$ by comparing
results from X-ray observations with the predictions of the Press-Schechter 
calculations depends on $\Omega_0$ differently from that from SZE
cluster `observations'. 

The total number density of SZE clusters for the fiducial model at 
$S_{\nu}^{lim}\approx 6.2\hbox { mJy}$ is $dN(z\ge 0)/d\Omega\approx
5.6 \hbox { deg}^{-2}$. With a $50\hbox { deg}^2$ survey, the total
number of SZE clusters is expected to be about $200$, and 
the standard deviation given by the Poisson statistics is then about $16.7$.
Thus the $3\sigma$ number density is $5.6\pm 1.0 \hbox { deg}^{-2}$. 
The corresponding $\sigma_8$ range is calculated, which can be well 
approximated by $\sigma_8=(0.794\pm 0.025)\Omega_0^{-0.13}$. In Fig. 7, we plot
the $\sigma_8-\Omega_0$ relations expected from both the X-ray observations 
(dashed lines)
and from our analysis on the SZE clusters (solid lines). Based on the current
X-ray cluster results and our proposed $50\hbox { deg}^2$ SZE cluster
survey, the fiducial $\Omega_0$ can be determined to 
$\Omega_0=0.3\pm 0.08$ or $\Omega_0=(1\pm 27\%)\times 0.3$.
From Fig. 7, it can be seen that the X-ray cluster-normalization 
constrains more tightly the value of $\Omega_0$ as it
is much more sensitive to $\Omega_0$. With future
X-ray cluster surveys such as XMM, the normalization can be 
determined to a much better degree. At a $2\%$ precision on 
$\sigma_8$ from X-ray surveys, $\Omega_0$ can be constrained
to $0.3\pm 0.04$ ($\sim 13\%$ precision) with the allowance of 
a $3\sigma$ deviation of SZE cluster counts.  
In comparison with the results of Haiman et al. (2000, Fig.7 in 
their paper) for the $w=-1$ case, their constraint on $\Omega_0$
is more stringent. But in their analyses, they normalize all models 
to give rise to the local cluster number density with $M\ge 10^{14}h^{-1}
\hbox { M}_{\odot}$, i.e., for each model, the normalization is
fixed. They then compare both the total number of SZE clusters
and the redshift distribution of a model with those of their fiducial one
to constrain the parameter regimes. Note that for the $w=-1$ case, 
the total number of SZE clusters plays the dominant role in constraining
$\Omega_0$. If we focus on the single middle dashed line 
in our Fig. 7, the $3\sigma$ determination of $\Omega_0$ from the
SZE total number counts is similar to that of Haiman et al. (2000).
Be aware however the different cosmological parameters, the normalizations
and the survey parameters used in their analyses and in our studies.
     
The normalization factor $\sigma_8$ can also be constrained at the
same time. In contrary to the determination of $\Omega_0$, the $\sigma_8$ 
value is constrained more tightly by the SZE cluster counts.
At $\Omega_0=0.3$, the $3\sigma$ determination
of $\sigma_8$ is $0.93\pm 0.03$ or $(1\pm 3.2\%)\times 0.93$ in 
comparison with $(1\pm 7.5\%)\times 0.93$ from the current X-ray studies.

We emphasize that although we chose $\Gamma=0.25$ in the above analysis,
calculations have been done for other $\Gamma$ values. It is found that 
the functional relation $\sigma_8\propto\Omega_0^{-0.13}$ is very 
insensitive to the $\Gamma$ value. Moreover the $S_{\nu}^{lim}$ value
has almost no effect on this functional relation.






\section{Summary}


In our analyses, we used $f_{ICM}=0.1$, $X=0.76$ and
$[d \hbox {ln} \rho_{gas}(r)/ d\hbox {ln} r]_{r_{vir}}=2$. To change these
parameters however, is equivalent to change the overall flux limit
$S_{\nu}^{lim}$. From the figures we showed, it is easy to see that
all our conclusions remain qualitatively unchanged for different choices
of these parameters. We have assumed the hydrodynamic equilibrium and
isothermality for the intracluster gas. The Press-Schechter
formalism has been adopted to calculate the cluster counts.

We studied the $r$ quantity for two types of cosmologies: the low-density
flat models and the low-density open models. Within the studied parameter
regimes $0.25\le \Omega_0 \le 0.35$ and $0.2\le \Gamma \le 0.3$,
the $r$ value for the two sets of cosmologies are well separated.
The flat model with $\Omega_0=0.3$ and $\Gamma=0.25$ and the open model
with the same $\Omega_0$ and $\Gamma$ can be differentiated at
the $3\sigma$ level for a $50 \hbox { deg}^2$ survey. Since we normalize
different models in a way such that they give rise to the same total number of
SZE clusters, our analyses naturally take into account the total number of
SZE clusters. For wider parameter ranges, the information from other
observations, such as X-ray or optical cluster surveys has to be
used to differentiate the two types of models.

There are other ways to determine the geometry of the universe.
Measurements on fluctuations of the CMB
radiation provide a clean test on this aspect (e.g., Hu 2000).
The horizon size of the
universe at decoupling separates large-scale and small-scale CMB
fluctuations. Large-scale fluctuations were outside the horizon when
photons escaped while small-scale perturbations were within the
horizon at decoupling and therefore sustained acoustic oscillations.
The position of the primary Doppler peak of the CMB fluctuation
power spectrum is determined by the angular size of the
horizon at decoupling. For a flat universe (low-density with a non-zero
cosmological constant or high density), the peak located at $l\approx 200$, and
the peak is shifted to smaller angular scale or higher $l$ for a low-density
open universe, where $l$ represents the two-dimensional angular wavenumber.
Observations have seen the rising and the declining of CMB
fluctuations around $l\sim 200$ (e.g., Miller et al. 1999, de Bernardis
et al. 2000), which constrains convincingly that
$\Omega_{tot}$ is close to $1$, where $\Omega_{tot}$ is the total
density parameter. Combining with other information, e.g., supernova
measurements (Perlmutter et al. 1999, Schmidt et al. 1998), leads to the
conclusion that we are living in a low-density flat universe.
If the results from SZE cluster surveys are in agreement with the CMB
results, we will be in a more solid
position to say that the universe is flat and the simple
structure formation theory we adopted here is reasonably correct.
On the other hand, any inconsistency between the SZE cluster results and
the results from CMB measurements would pose challenges to our
understanding of the universe. For example, alternate structure
formation theories, such as non-Gaussian initial fluctuation models,
may need to be considered.

Should the universe be flat, SZE cluster surveys can also
provide constraints on cosmological parameters.
We studied $\sigma_8(\Omega_0)$ inferred from the
total SZE cluster counts. A functional relation $\sigma_8\propto\Omega^{-0.13}$
is found. Combined with the current X-ray cluster-normalized $\sigma_8$,
the parameters can be determined (take $\Omega_0=0.3$ and $\sigma_8=0.93$
as the central values) to $\Omega_0=(1\pm 27\%) \times 0.3$ and
$\sigma_8=(1\pm 3.2\%)\times 0.93$
for the SZE cluster counts confined to the $3\sigma$ level in a
$50\hbox { deg}^2$ survey. Note that these constrains are from the
total number of clusters only, and no redshift information is needed.



\acknowledgments

We thank the referee for constructive comments and H. Liang for 
valuable discussions. This research was supported in part by 
grants NSC89-2112-M002-037 and NSC89-2816-M-001-0006-6 from the 
National Science Council, R.O.C.

\clearpage



\figcaption[fig1.ps]{
Redshift distributions of SZE clusters with $S_{\nu}^{lim}
\approx 6.2 \hbox { mJy}$. The solid line is for $\tau$CDM model
with $\Omega_0=1$, $h=0.5$, $\Gamma=0.25$ and $\sigma_8=0.52$.
The dash-triple-dotted line is for the SCDM model with
$\Omega_0=1$, $h=0.5$, $\Gamma=0.5$ and $\sigma_8=0.52$.
The dash-dotted line is for the open CDM model with $\Omega_0=0.3$,
$\Gamma=\Omega_0 h=0.25$ and $\sigma_8=0.87$. The dotted line is
for the $\Lambda$CDM model with $\Omega_0=0.3$, $\Omega_{\Lambda}=0.7$,
$\Gamma=\Omega_0 h=0.25$ and $\sigma_8=0.93$.
\label{fig1}}

\figcaption[fig2.ps]{
The ratio $r$ against the flux limit $S_{\nu}^{lim}$.
The solid line is for the fiducial model with $\Omega_0=0.3$,
$\Omega_{\Lambda}=0.7$, $\Gamma=\Omega_0 h=0.25$ and $\sigma_8=0.93$.
The upper dotted line is for the open model with $\Omega_0=0.35$,
$\Gamma=\Omega_0h=0.3$, and $\sigma_8=0.785$, and the lower
dotted line is for the open model with $\Omega_0=0.25$,
$\Gamma=\Omega_0h=0.2$, and $\sigma_8=0.895$. The three dash-dotted lines
are for $\Omega_0=0.3$ open models with $\Gamma=\Omega_0h=0.3,
0.25$ and $0.2$, from top to bottom respectively, and the corresponding
$\sigma_8=0.79, 0.835$ and $0.89$.
\label{fig2}}

\figcaption[fig3.ps]{
$r$ versus $S_{\nu}^{lim}$. The solid line is for
the $\Lambda$CDM model with $\Omega_0=0.25$, $\Omega_{\Lambda}=0.75$,
$\Gamma=\Omega_0h=0.2$, and $\sigma_8=1.01$ (Model 1).
The dotted line is for the open model with $\Omega_0=0.35$,
$\Gamma=\Omega_0h=0.3$ and $\sigma_8=0.785$ (Model 2). The three dashed lines
are for the open models with $\Omega_0=0.25$, and from top to bottom
$\Gamma=\Omega_0h=0.3, 0.25$ and $0.2$, respectively (Model 3, 4, 5).
The respective $\sigma_8$ for the open models with $\Omega_0=0.25$ are
$0.795, 0.84$ and $0.895$.
\label{fig3}}

\figcaption[fig4.ps]{
$r$ versus $S_{\nu}^{lim}$ for $\Omega_0=0.3$ and
$\Omega_{\Lambda}=0.7$. The solid line is for $\Gamma=\Omega_0h=0.25$,
and $\sigma_8=0.93$. The dotted line is for $\Gamma=\Omega_0h=0.3$
and $\sigma_8=0.88$. The dashed line is for $\Gamma=\Omega_0h=0.2$
and $\sigma_8=0.99$.
\label{fig4}}

\figcaption[fig5.ps]{
$r$ versus $S_{\nu}^{lim}$ for $\Lambda$CDM models with
$\Gamma=\Omega_0h=0.25$. The solid line is for $\Omega_0=0.3$,
$\Omega_{\Lambda}=0.7$, and $\sigma_8=0.93$. The dotted line is for
$\Omega_0=0.35$, $\Omega_{\Lambda}=0.65$, and $\sigma_8=0.91$.
The dashed line is for $\Omega_0=0.25$, $\Omega_{\Lambda}=0.75$
and $\sigma_8=0.95$.
\label{fig5}}

\figcaption[fig6.ps]{
SZE cluster-normalized $\sigma_8$ as a function of $\Omega_0$
for flat universes with a non-zero cosmological constant.
The stars are the numerical results calculated by requiring that models
with different $\Omega_0$ contain the same number of SZE clusters
at $S_{\nu}^{lim}\approx 6.2\hbox { mJy}$ as that of the fiducial model.
The fitting relation $\sigma_8=0.794\Omega_0^{-0.13}$ is shown as the
solid line.
\label{fig6}}

\figcaption[fig7.ps]{
$\sigma_8$ versus $\Omega_0$ for flat universes. The solid
lines are from the SZE cluster counts, and from top to bottom,
the respective $\sigma_8$ are $\sigma_8=0.819\Omega_0^{-0.13}$,
$\sigma_8=0.794\Omega_0^{-0.13}$ and $\sigma_8=0.769\Omega_0^{-0.13}$.
The dashed lines are from X-ray temperature limited cluster counts,
and $\sigma_8=0.56\Omega_0^{-0.52+0.13\Omega_0}$,
$0.52\Omega_0^{-0.52+0.13\Omega_0}$, and $0.48\Omega_0^{-0.52+0.13\Omega_0}$,
respectively, from top to bottom.
\label{fig7}}

\clearpage






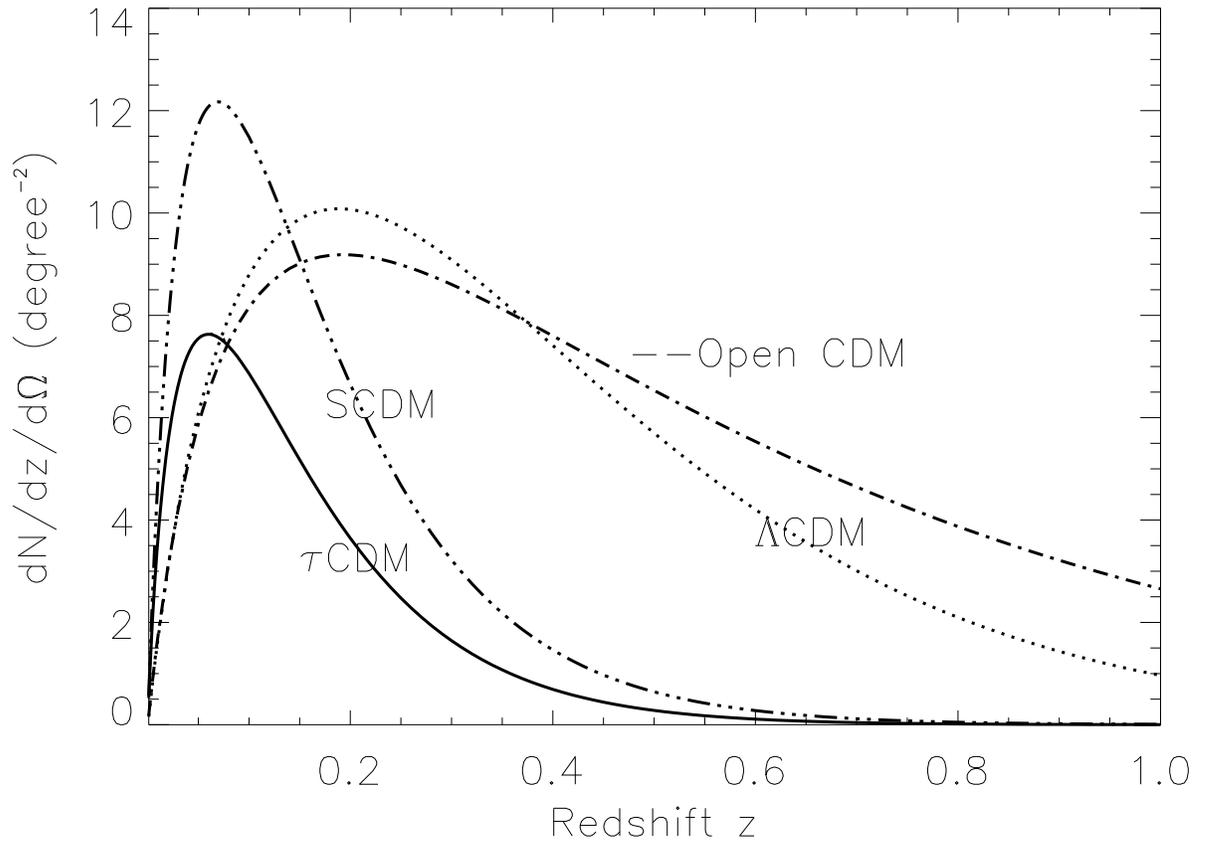
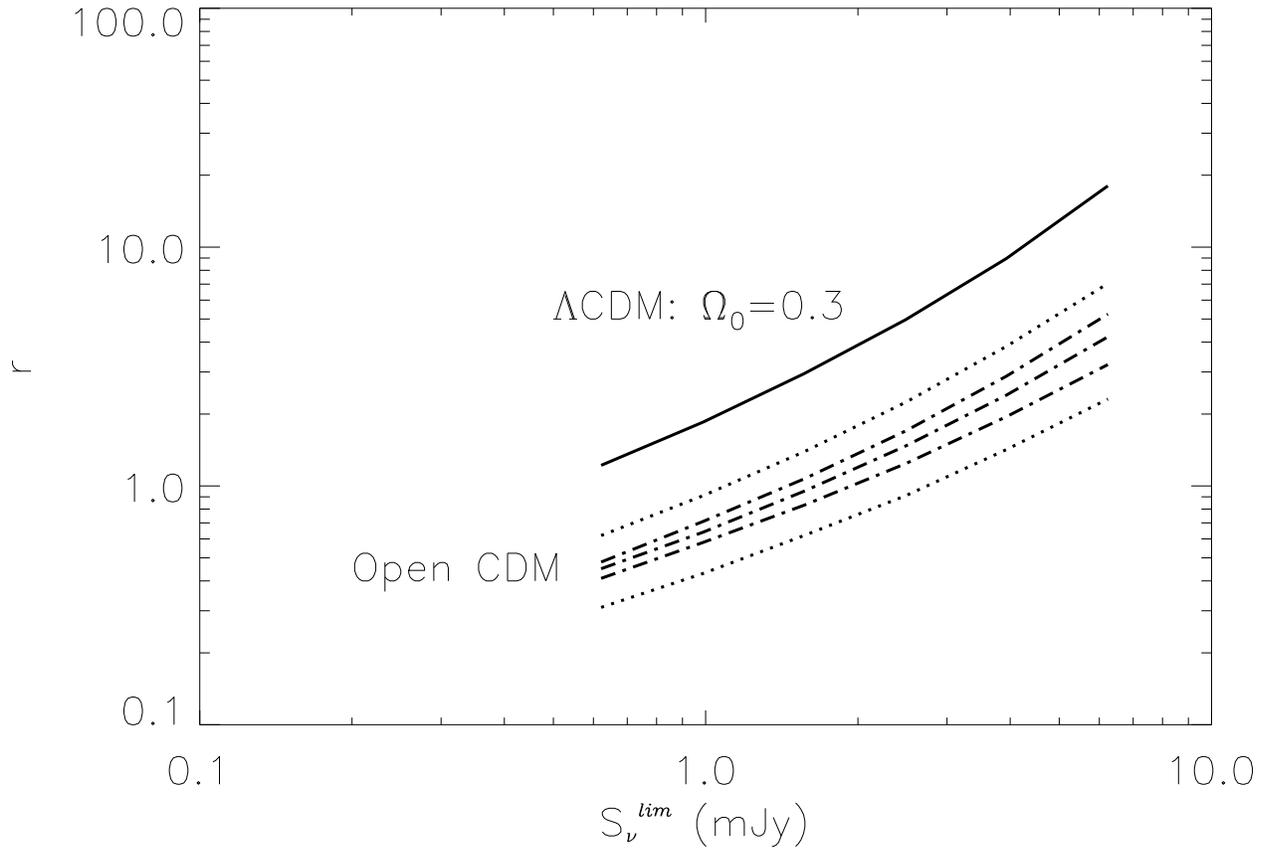
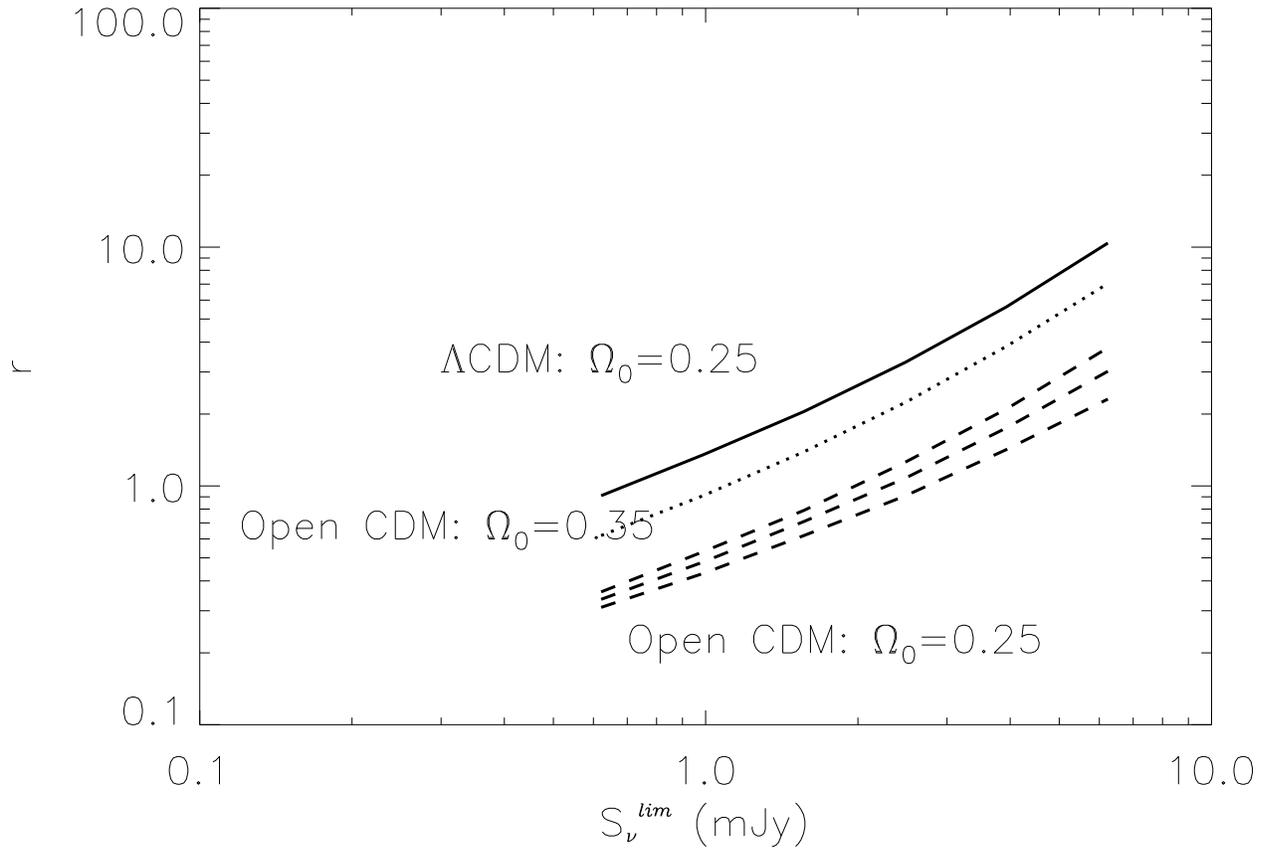
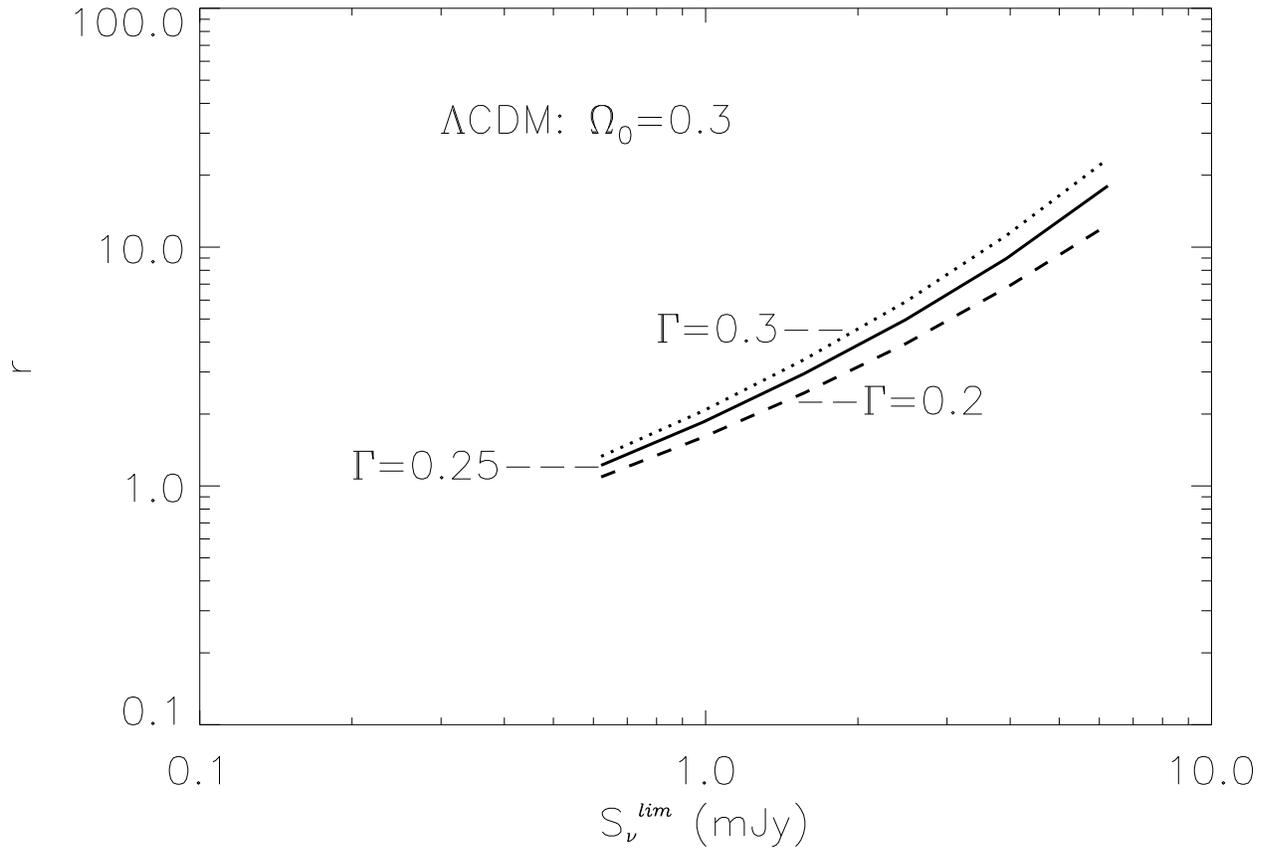
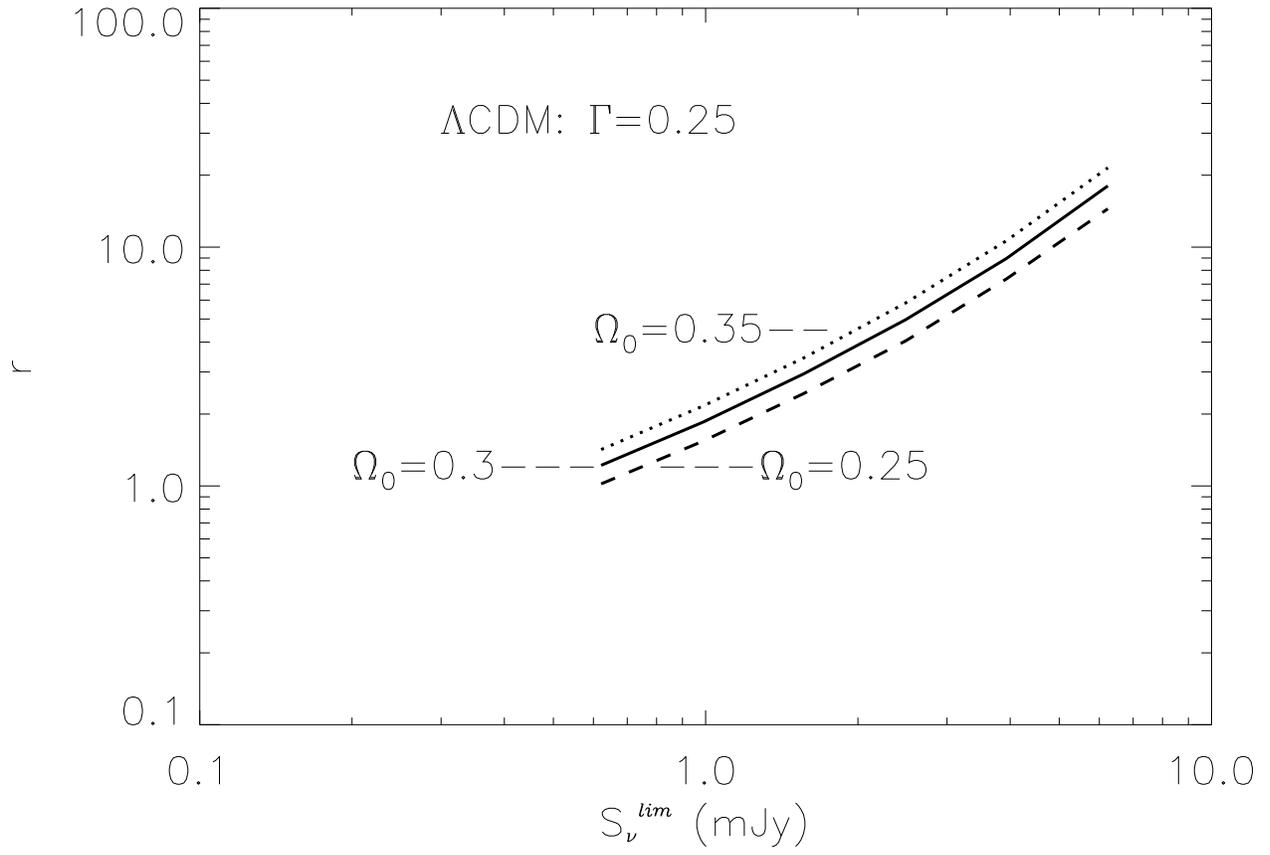
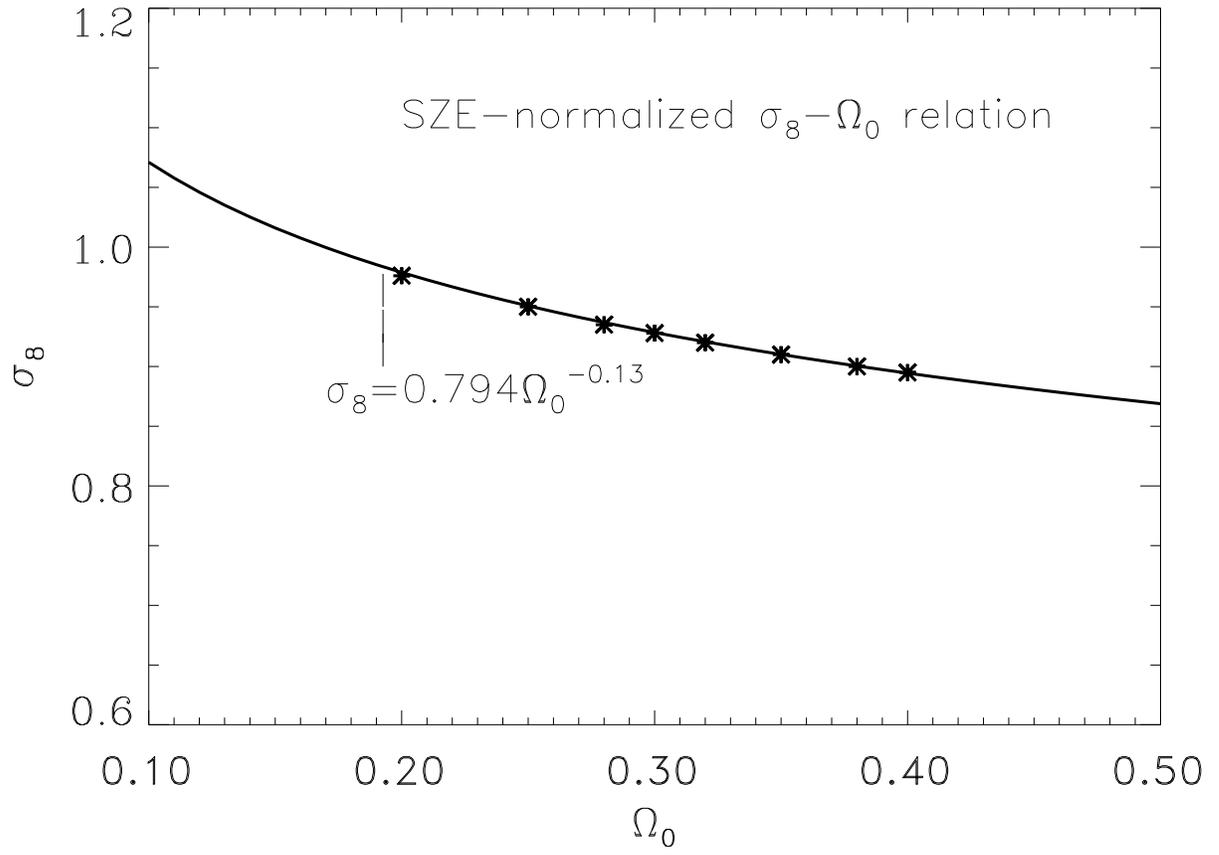
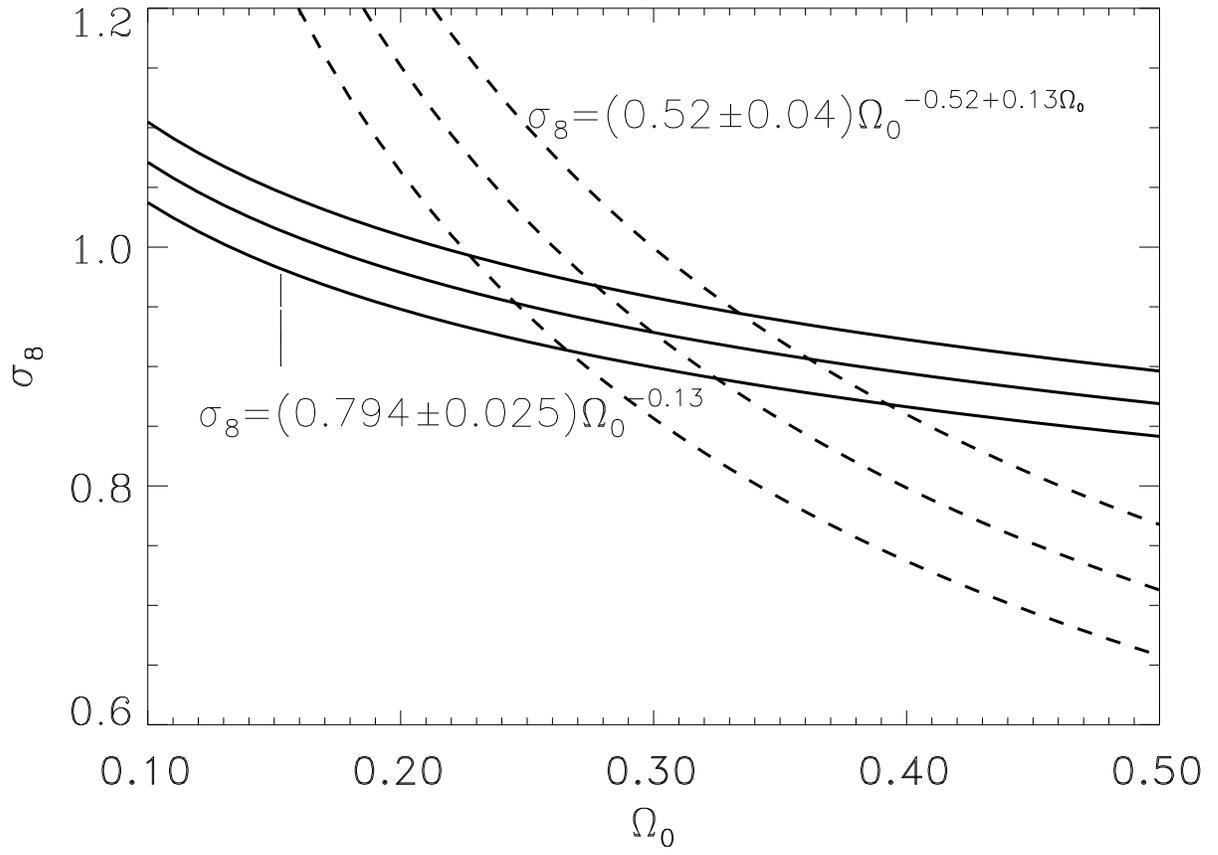
\end{document}